\documentclass[10pt,aps,prl,raggedbottom,longbibliography,reprint,citeautoscript,letterpaper,notitlepage,floatfix]{revtex4-2} 
\usepackage[usenames,dvipsnames]{color}
\usepackage{graphicx,microtype}
\usepackage{dcolumn}
\usepackage[bookmarks=false,colorlinks]{hyperref}
\hypersetup{linkcolor=magenta,citecolor=MidnightBlue,filecolor=Plum,urlcolor=MidnightBlue}
\usepackage[all]{hypcap} 
\usepackage{lmodern}

\newcommand{\supf}{\textcolor{DarkOrchid}{Fig.}\,S}



\graphicspath{{./Figures}}

\begin{document}

\title{Suppressed paramagnetism in amorphous Ta$_2$O$_{5-x}$ oxides and its link to superconducting qubit performance}

\author{P.\ Graham Pritchard}
\email{patrickpritchard2025@u.northwestern.edu}
\affiliation{Department\;of\;Materials\;Science\;and\;Engineering, Northwestern University, Evanston, Illinois  60208, USA}

\author{James M.\ Rondinelli}%
\email{jrondinelli@northwestern.edu}
\affiliation{Department\;of\;Materials\;Science\;and\;Engineering, Northwestern University, Evanston, Illinois  60208, USA}
\date{\today}

\begin{abstract}
Reduced transmon qubit $T_1$ coherence times have been linked to the amorphous oxide layers formed by  thin film capacitors during processing.
Because Ta or Ta capped Nb capacitors exhibit overall superior qubit performance to those fabricated with Nb capacitors, it has been hypothesized that the amorphous, non-stoichiometric Ta$_2$O$_{5-x}$ oxide is less lossy than its Nb$_2$O$_{5-x}$ counterpart.
The origins of what makes amorphous Ta$_2$O$_{5-x}$ less susceptible to accepted decoherence channels is unknown. 
Here we establish the microscopic features of amorphous Nb$_2$O$_{5-x}$ and Ta$_2$O$_{5-x}$ using a combination of \textit{ab initio} molecular dynamics and density functional theory calculations. 
Our simulations establish that oxygen deficiency is less likely to occur in amorphous Ta$_2$O$_{5-x}$ than in Nb$_2$O$_{5-x}$ for $0\le x \le 0.25$ and that at a given level of oxygen deficiency the formation of metal Ta-Ta bonds is enhanced.
These bonds, which are accommodated by structural flaws in the amorphous network, capture electrons better than in amorphous Nb$_2$O$_{5-x}$.
These thermochemical differences quench or highly suppress magnetic moments in amorphous Ta$_2$O$_{5-x}$ and eliminate a potential source of quasiparticles and magnetic flux noise. 
Our calculations also show that hyperfine couplings between Nb nuclei and local magnetic moments in Nb$_2$O$_{5-x}$ could form ``two-level systems'' (TLS) or ``two-level fluctuators'' (TLF) with energy splittings of 100-1000 MHz or higher. This reveals a new TLS mechanism in amorphous Nb$_2$O$_{5-x}$ oxide layers that is unlikely in Ta$_2$O$_{5-x}$.
Our work provides fundamental understanding of the materials chemistry and limitations imposed by native oxides of superconducting qubits, which can be used to guide materials selection and  processing.
\end{abstract}
\maketitle
\sloppy


\paragraph{Introduction.}
Two-dimensional superconducting transmon qubits \cite{Koch2007} are a leading platform for quantum computing with state-of-the-art designs possessing peak $T_1$ coherence times in excess of 0.3-0.5 ms \cite{Place2021, Wang2022, Bal2023}, single qubit gates errors below $10^{-3}$\cite{Acharya2023} to $10^{-4}$ \cite{Li2023}, and two-qubit gate errors below $10^{-2}$ \cite{Acharya2023}. Improvements to these performance metrics are desired to implement robust and scalable fault-tolerant quantum computers. When optimized gate pulses are implemented on transmon qubits \cite{Li2023}, further improvements to single qubit error rates are predominately limited by decoherence mechanisms rather than further optimization of gate pulses. Thus, understanding and mitigating single qubit decoherence mechanisms remains a critical hurdle to advance future quantum computing technologies. 

Coherence times of superconducting qubits are thought to be limited by the presence of atomic scale microstructural imperfections, which produce two-level system defects (TLS) and/or introduce pair breaking mechanisms that produce quasiparticles (QP) \cite{Muller2019, Murray2021}. This is founded on a combination of experimental and theoretical efforts, which have correlated phenomenological predictions of TLS and QP models with observed trends in the temperature, frequency, noise, and power dependence of qubit coherence times and proxy devices such as microwave resonators \cite{Gao2008, Lisenfeld2015, Faoro2015, Serniak2018, Graaf2020, Crowley2023}. Despite this understanding, the physical structural description and origins of the dominant defects leading to decoherence remain elusive \cite{Muller2019}, because of the heterogeneous nature of qubit materials and interfaces, structure sensitivity to processing, and complexity in direct deterministic probes of defect-$T_1$ relationships.

While high coherence ($T_1 > 0.1$ ms) transmon qubits have been fabricated with either Al, Nb, or Ta capacitor pads in combination with an Al/AlO$_x$/Al Josephson junction \cite{Ganjam2023,Nersisyan2019,Place2021,Wang2022}, transmon qubits with Ta capacitor pads or Ta capped Nb pads show consistently improved  $T_1$ times compared to qubits fabricated with Nb \cite{Place2021, Bal2023} or Al pads \cite{Ganjam2023} after maintaining nominally the same geometry and fabrication processes. As the superconducting transition temperature of Ta films ($T_c\sim 4.3$\,K) \cite{Face1987,Place2021} is less than that of Nb films ($\sim$ 9.3 K) \cite{Tanatar2022}, the improved performance observed with Ta pads is not attributable to a reduction in intrinsic QP losses. Rather, it may reside in its native or process-oxide layer.

Ta, Nb, and Al metals films all form amorphous process-oxide layers \cite{Altoe2022, Bal2023}. These oxides are expected to host polarized TLS defects which couple with the electric field of the qubit \cite{Muller2019}. Indeed, $T_1$ times of transmon qubits with Al capacitor pads have been shown to inversely correlate with the electric field participation ratio of the metal-air, metal-substrate, and substrate-air interfaces, which indicates a dielectric loss mechanism impairs transmon $T_1$ \cite{Wang2015}. More recently, an $\sim$5 nm amorphous process oxide layer (NbO$_x$) has been reported to reduce the single-photon quality factor of superconducting Nb thin film resonators \cite{Altoe2022}. 
Removal of this oxide layer, while limiting regrowth to $\sim$1 nm, was correlated with a more than 2-fold improvement in the resonator quality factor. Decomposition of resonator losses into TLS and high-power components has shown that more than 50\% of this improvement is due to non-TLS losses. Thus, dielectric losses arising from TLS are \textit{not} the sole source of loss within the amorphous oxides. In fact, oxygen-deficient Nb pentoxides host paramagnetic moments \cite{Cava1991,Ruscher1991,Fang2015}. Such moments in the boundary layer of a superconductor can produce impurity states via the Shiba mechanism \cite{Shiba1968,Sheridan2021}, potentially explaining some of the observed high-power losses in Nb resonators. 

The extent to which these loss mechanisms also pertain to amorphous Ta oxides is unknown. A microscopic description is necessary to understand why Ta comprising qubits outperform traditional Nb-based circuits.
To that end, we use a combination of \textit{ab initio} molecular dynamics and density functional theory calculations to show that amorphous Ta oxides are less prone to oxygen deficiencies, consistent with experiment, due to their strong oxygen affinity. We then identify and describe the materials origin and suppression of paramagnetism in amorphous Ta and Nb oxides films, which arises from a difference in the tendency for electrons to fill band states rather than states within the electronic band gap.
This behavior explains the increased potential for quasiparticle or flux noise in Nb oxides compared to Ta oxides.
We then introduce a new class of TLS defects in amorphous Nb oxides, which arise from hyperfine couplings between Nb nuclei and local moments on Nb sites, that are a consequence of the persistent paramagnetism.

\paragraph{Oxygen Deficiency.} Motivated by recent experiments   that indicate Ta oxide thin films are more robust to oxygen loss than equivalently processed Nb oxide  films \cite{Bal2023}, we first address the susceptibility to oxygen substoichiometry $x$ in the amorphous state.
This establishes a base expectation for the range of oxygen deficiency in the films, which affects the magnetic defect density. 
We compute the formation energy of the amorphous Ta$_2$O$_{5-x}$ and Nb$_2$O$_{5-x}$ oxides using a combination of $\textit{ab initio}$ molecular dynamics (AIMD), to generate amorphous oxide structures across the compositional range $0\le x\le0.25$, and spin-polarized density functional theory (DFT) calculations. 
Details of the AIMD %
\footnote{Our AIMD structure generation methodology closely follows that reported in \cite{Bondi2013}, which successfully produced amorphous Ta$_2$O$_{5-x}$ structures consistent with experimentally inferred pair distribution functions \cite{Alderman2018}.}
and electronic structure calculations are given in the Supplementary Materials (SM) \cite{supp} and in Ref.\ \footnote{As the proper functional choice for simulating the magnetic behavior of Nb/Ta oxides is ambiguous in the literature, we evaluate the influence of functional choice on magnetism in our amorphous structures in deatil in the SM \cite{supp}. 
Like Nb crystallographic shear structures \cite{Kocer2019}, the electronic structure of magnetic states is strongly dependent on functional selection (\supf6-12). We find that the PBE functional is insufficient to produce a $d-d$ gap between filled and unfilled `conduction band' states, while all other functionals predict such a gap. Thus, higher level functional are needed in order to match the expected magnetic behavior.
Atomic relaxations of the AIMD generated structures and spin-polarized calculations were repeated using the PBE$+U$, r$^2$SCAN, and HSE06 functionals. PBE$+U$ functional calculations were performed with a $U=4$\,eV specified on transition metal sites (Nb/Ta). Owing to their additional computational cost, the HSE06 and r$^2$SCAN functionals were evaluated at a single fixed oxygen concentration ($x=0.25$).}.
The formation energy $E_f = E(\mathrm{M}_2\mathrm{O}_{5-x}) + N_\mathrm{O} \frac{1}{2}\mu_{\mathrm{O}_2} - E_{avg}\left(\mathrm{M}_2\mathrm{O}_5\right),$
where $E(\mathrm{M}_2\mathrm{O}_{5-x})$ is the total energy of an oxygen-deficient amorphous structure with M = (Ta, Nb), $E_{avg}(\mathrm{M}_2\mathrm{O}_5)$ is the average total energy of 7
stoichiometric Ta$_2$O$_{5}$ or Nb$_2$O$_{5}$ structures, $\mu_{\mathrm{O}_2}$ is the PBE total energy of the spin-polarized 
$\mathrm{O}_2$ molecule, and $N_\mathrm{O}$ is the number of oxygen atoms removed to achieve the target stoichiometry. (See the SM for additional methodological details \cite{supp}.)
\supf1 shows a larger increase in the formation energy per removed oxygen atom for  Ta$_2$O$_{5-x}$ than Nb$_2$O$_{5-x}$ (4.5 eV/atom and 3.5 eV/atom, respectively) over the $0\le x\le0.25$ compositional range, which corresponds to the oxygen deficiencies known to produce paramagnetic susceptibilities in crystalline Nb$_2$O$_{5-x}$ \cite{Cava1991, Ruscher1991}.
This result indicates that amorphous Ta$_2$O$_{5-x}$ is less likely to be oxygen deficient, consistent with the higher oxygen affinity of Ta metal compared to Nb metal \cite{Machlin1968}.  

\begin{figure}
\centering
   \includegraphics[width=0.48\textwidth]{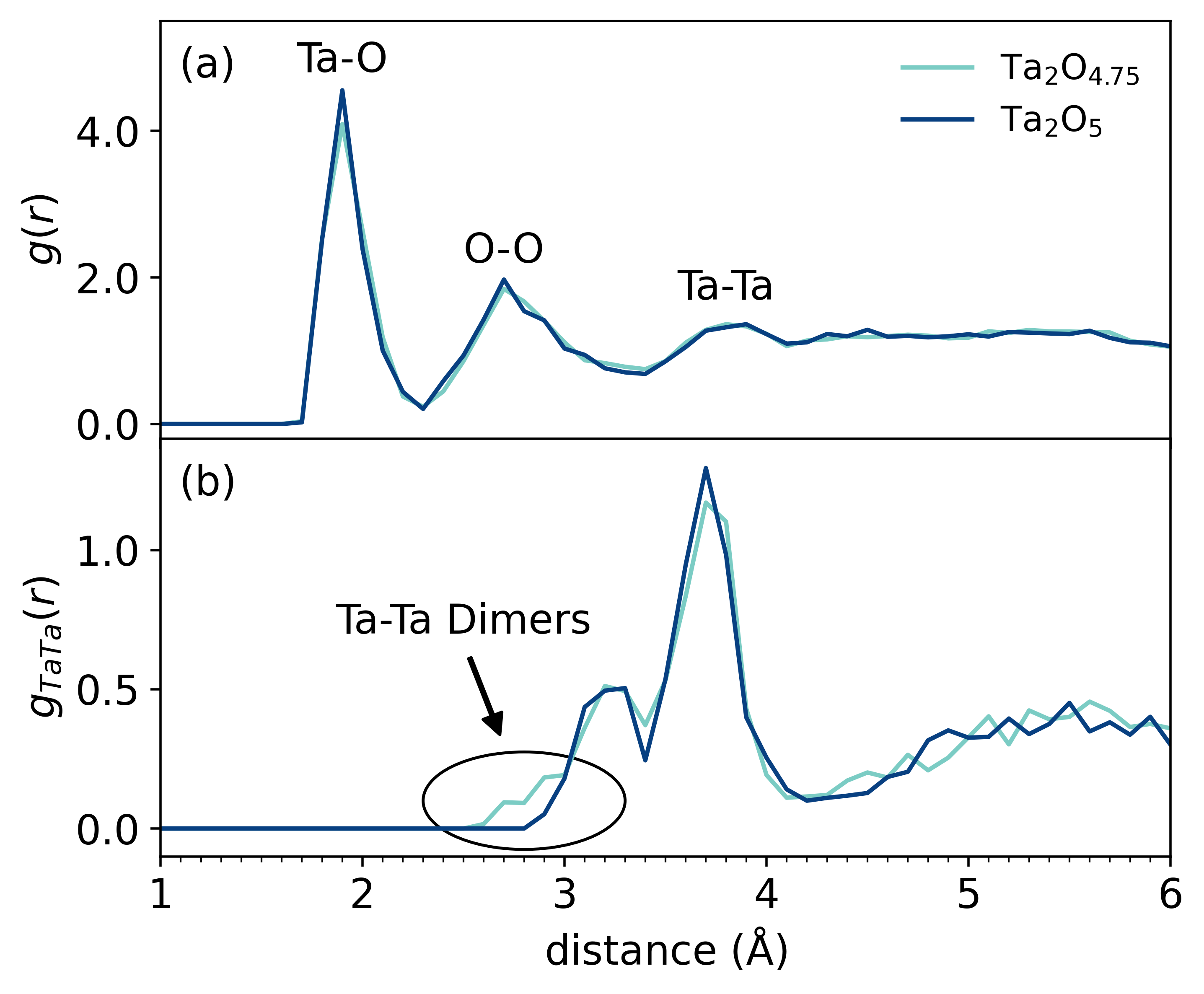} 
   \caption{(a) Total and (b) Ta-Ta DFT-PBE simulated mean pair distribution functions (PDF) of amorphous Ta$_2$O$_{4.75}$ and Ta$_2$O$_{5}$.  The $g_\mathrm{TaTa}(r)$ is normalized by the ratio of the number of Ta atoms to the total number of atoms in the full structure.}
   \label{fig:pdf}
\end{figure}

\paragraph{Local Structure.} To understand how the oxygen deficency is accommodated in the amorphous oxides, we computed the pair distribution functions (PDFs) for each relaxed amorphous structure (in a 3$\times$3$\times$3 supercell configuration) using the \texttt{rdfpy} python library \cite{rdfpy}. 
\autoref{fig:pdf}a presents the trend in the calculated PDFs as a function of oxygen content for the PBE functional. For clarity, each PDF represents the mean of the 7 PDFs computed at each oxygen concentration. We used an identical procedure to compute partial Ta-Ta PDFs, $g_\mathrm{TaTa}(r)$ (\autoref{fig:pdf}b). We note that our PDFs for stoichiometric Ta$_2$O$_5$ are consistent with PDFs derived from prior AIMD  simulations \cite{Bondi2013} and experimentally derived PDFs for amorphous Ta$_2$O$_5$ \cite{Alderman2018}.

We identify three nearest-neighbor atoms pairs: Ta-O (1.9\,\AA), O-O (2.7\,\AA), and  Ta-Ta (3.8\,\AA). Other than a reduction in the Ta-O peak, there is no discernable impact of the oxygen concentration on these mean PDFs at the resolution presented in \autoref{fig:pdf}a. 
A nearest-neighbor analysis indicates that Ta is primarily six- and five-coordinate and inspection of the amorphous structures shows irregular TaO$_6$ and TaO$_5$ polyhedra.
Notably in the amorphous structures, we find two distinct peaks in the mean Ta-Ta PDFs at all oxygen concentrations: the first at $\sim$3.2 \AA\ and the second at $\sim$3.7 \AA. The former corresponds closely to the second nearest-neighbor bond distance in body-centered cubic (BCC) Ta (3.3 \AA) \cite{Edwards1951}. The latter peak is due to O-bridged Ta atoms. As the oxygen concentration decreases $x=0\rightarrow0.25$, an additional feature appears between 2.6-3.0\,\AA\ and bounds the first nearest-neighbor distance of 2.86\,\AA\ in BCC Ta and is below twice the value of the covalent radius of Ta, indicative of increased metal-metal bonding in the amorphous oxide.

\begin{figure}
\centering
   \includegraphics[width=0.48\textwidth]{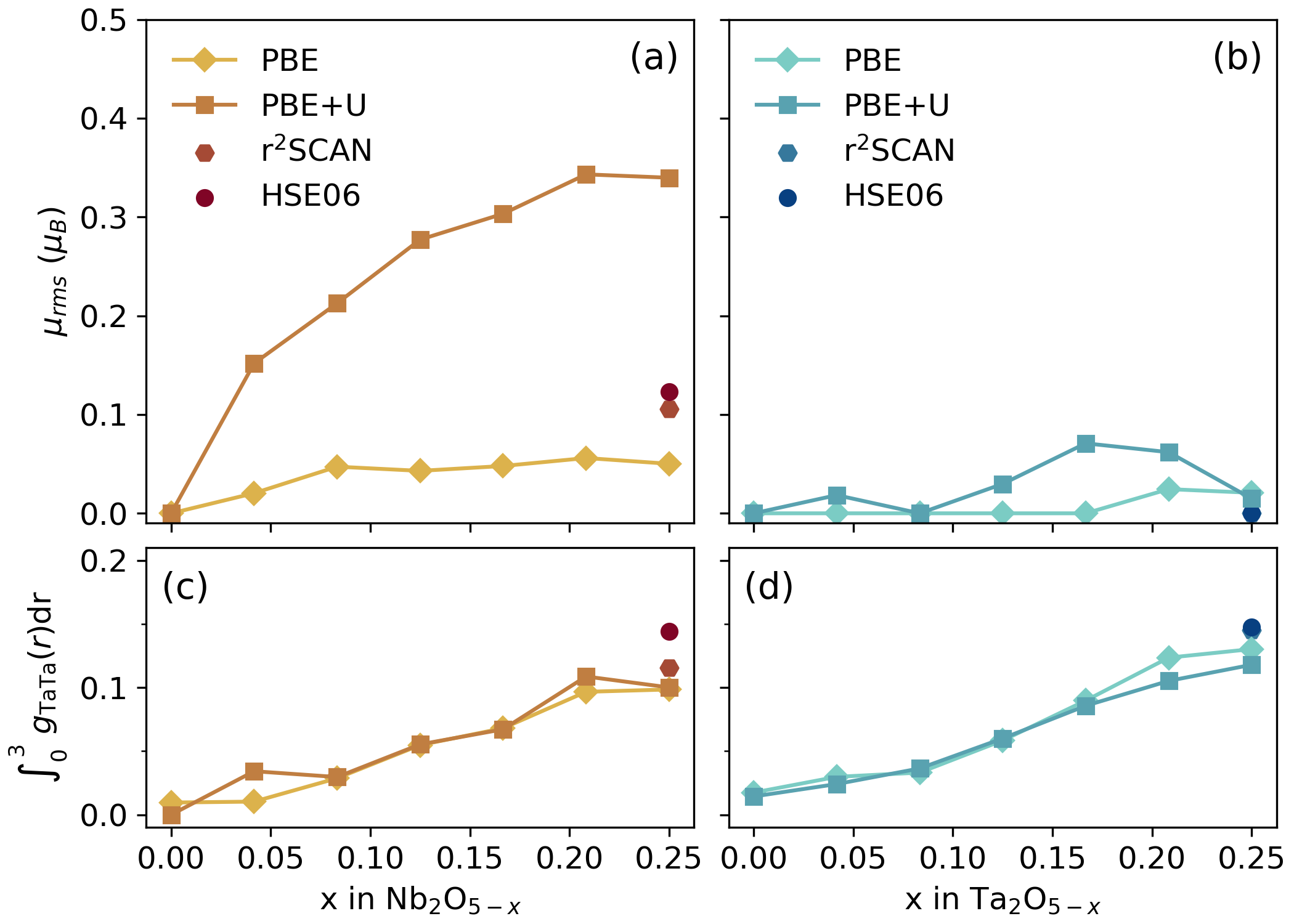} 
   \caption{Root-mean-square of magnetic moments hosted on transition metal sites in (a) Nb$_2$O$_{5-x}$ and (b) Ta$_2$O$_{5-x}$ from different DFT functionals. Integrated metal-metal PDFs in the range $r=0-3$\,\AA\ for (c) Nb$_2$O$_{5-x}$ and (d) Ta$_2$O$_{5-x}$.} 
   \label{fig:momentstats}
\end{figure}

\paragraph{Magnetism.}
The modified metal-oxygen bond network and shortened Ta-Ta distances in the amorphous oxygen deficient state allows for the formation of Ta-Ta dimers and Ta $N$-mers (see below).
These structural flaws in the glass network tune how electrons fill the magnetic states at the band edges and the primarily non-magnetic states within the band gap.
We evaluate the relative stability of a magnetic configuration for each amorphous oxide by calculating the spin-polarization energy, $\Delta E_\mu=E_\mu-E_\mathrm{NM}$, as the difference between the total energy of a relaxed spin-polarized simulation, $E_\mu$, and the total energy of the corresponding relaxed non-magnetic (NM) simulation, $E_\mathrm{NM}$. A spin-polarized solution is energetically preferred for $\Delta E_\mu < 0$. We find that the number of energetically stable magnetic configurations and the simulated spin-polarization energy are strongly dependent on chemistry and functional (\supf4 and \supf5 \cite{supp}). 
For Nb$_2$O$_{5-x}$ at the PBE level, 21 out of 42 oxygen-deficient structures hosted stable moments and $\Delta E_\mu$ ranged between -140 meV to +16 meV with an average value $\langle\Delta {E}_\mu\rangle =-31$ meV. At the PBE$+U=4$\,eV for the Nb $4d$ orbitals, the number of structures hosting stable moments increased to 41 out of 42 and the range of $\Delta E_\mu$ greatly increased: -2.6 eV to +81 meV with $\langle\Delta {E}_\mu\rangle =-790$\,meV. 
For Ta$_2$O$_{5-x}$ only 2 out of 42 structures at the PBE level and 9 out of 42 structures at the PBE$+U$ level hosted stable moments. In addition, the corresponding spin polarization energies ranged from $-36\le \Delta E_\mu \le -10$\, meV ($\langle\Delta {E}_\mu\rangle =-22$\,meV) and $-330 \le \Delta E_\mu \le 460$\,meV ($\langle\Delta {E}_\mu\rangle =-23$\,meV), respectively. 
Thus, even when magnetic solutions exist for Ta oxides, the energetic stability is reduced. 
While most Nb$_2$O$_{5-x}$ structures exhibit stable moments and  ordered configurations, e.g., antiferromagnetic (AFM), ferrimagnetic, and ferromagnetic (FM) regardless of functional, 
few Ta$_2$O$_{5-x}$ structures host any local moments.
Physically, the local moment formation can be understood as resulting from the filling of magnetic band states which are present due to lack of a sufficient number of non-magnetic gap states. As $\Delta E_\mu$ and the degree of ferromagnetism are not strongly correlated (see \supf3 and \supf5), we infer that the intrinsic spin-spin coupling is weak--consistent with the low-temperature paramagnetism of Nb crystallographic shear structures \cite{Cava1991}.
To ensure an accurate description of the electronic and magnetic interactions, we further employed the regularized-restored strongly constrained and appropriately normed (r$^2$SCAN) \cite{Furness2020, Devi_2022} and Heyd-Scuseria-Ernzerhof hybrid (HSE06) functionals \cite{Gopal_2017}. Here we found that with r$^2$SCAN, 5 out of 7 Nb$_2$O$_{5-x}$  structures hosted stable moments and $\Delta E_\mu$ ranged from -330 meV to -10 meV with $\langle\Delta {E}_\mu\rangle =-160$\,meV. With HSE06, 5 out of 7 structures also hosted stable moments and $\Delta E_\mu$ ranged from -380 meV to -9 meV with $\langle\Delta {E}_\mu\rangle =-260$\,meV. In contrast, no amorphous Ta$_2$O$_{5-x}$ structures hosted a stable magnetic configuration with either r$^2$SCAN or  HSE06.
Next, we evaluate the root-mean-square of the magnetic moments, $\mu_{rms}$, at a fixed level of oxygen deficiency by compiling the magnetic moments present on each transition metal site after relaxation for each of the AFM and FM initialized simulations (including those for which no magnetic solution is present). If the magnetic solution was unstable ($\Delta E_\mu > 0$), magnetic moments were set to zero for that simulation. \autoref{fig:momentstats} shows that $\mu_{rms}$ for Nb$_2$O$_{5-x}$ increases gradually from $x=0$ to $x=1/12$ and remains essentially constant at higher levels of oxygen deficiency at the PBE level.
With the PBE$+U$ functional, $\mu_{rms}$ increases monotonically with increasing oxygen deficiency. $\mu_{rms}$ is 110\% and 145\%  larger at the r$^2$SCAN and HSE06 levels compared to PBE, yet still three fold less than PBE$+U$. In contrast,  $\mu_{rms}=0$ for Ta$_2$O$_{5-x}$ structures below $ x=0.166 $ and remains lower than $\mu_{rms}$ for Nb$_2$O$_{5-x}$ structures for larger $x$ with the PBE functional. 
Although the PBE$+U$ $\mu_{rms}$ values are larger than the PBE values, they remain substantially suppressed in Ta$_2$O$_{5-x}$ compared to Nb$_2$O$_{5-x}$ and do not increase monotonically.
These results indicate that amorphous Ta$_2$O$_{5-x}$ oxides have a much weaker propensity for hosting magnetism, and support the use of the higher fidelity  r$^2$SCAN and HSE06 calculations giving $\mu_{rms}=0$.

\paragraph{Hyperfine States.}
Atomic Nb has a large nuclear spin ($I=9/2$) and an energy splitting of 9,341 MHz between the hyperfine sublevels of its $^6D_{1/2}$ ground state \cite{Buttgenbach1975}. Correspondingly, resonant transitions between its hyperfine sublevels are active in the microwave frequency range of transmon qubits (i.e., 1-10 GHz) \cite{Roth2023}. To evaluate whether hyperfine splittings may still be of a relevant magnitude in Nb oxides, we calculated the hyperfine tensor $A_{ij}$ and the correspondingly hyperfine splittings for each Nb site hosting a local moment \cite{supp}. 

\begin{figure}
\centering
   \includegraphics[width=0.48\textwidth]{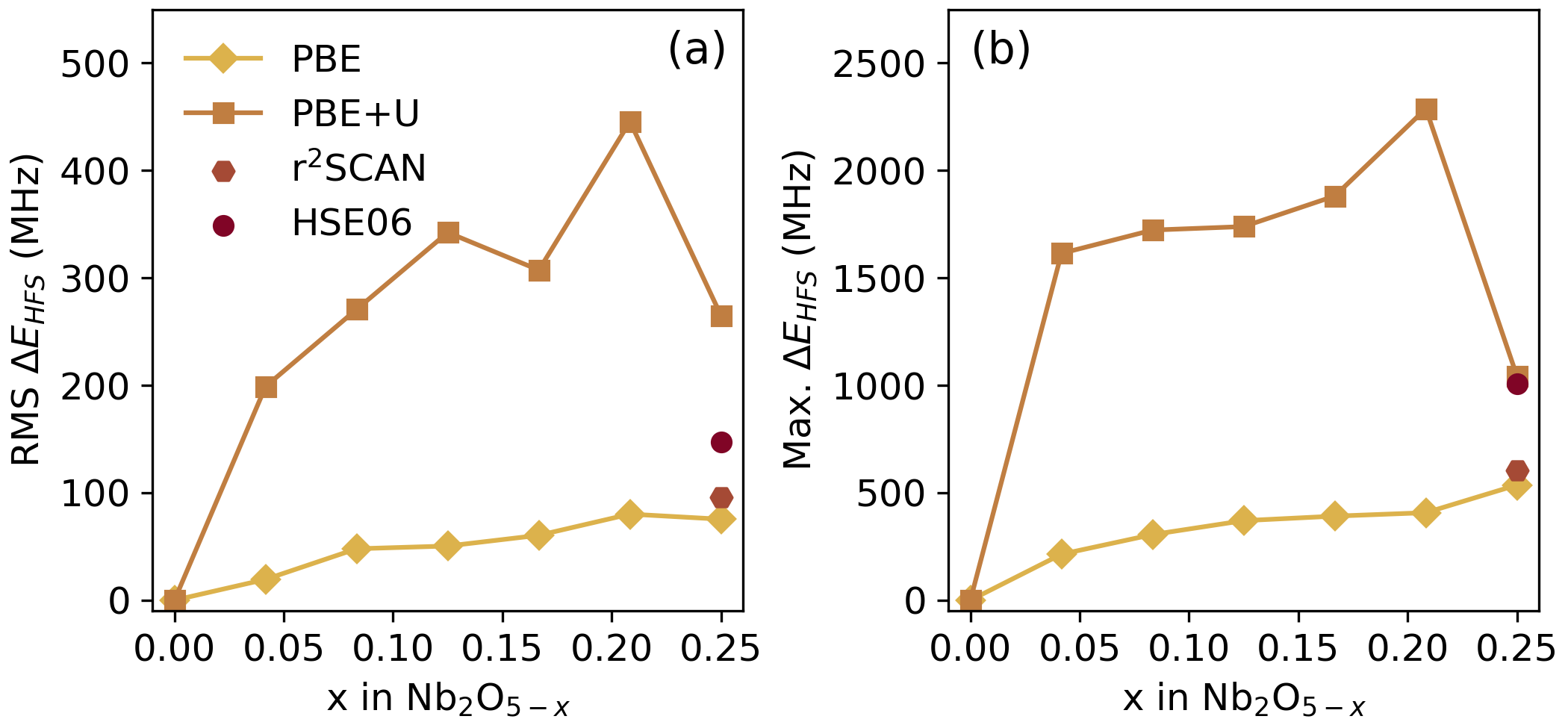} 
   \caption{(a) Root-mean-square (RMS) and (b) maximum hyperfine energy splittings ($\Delta E_{HFS}$) in amorphous Nb$_2$O$_{5-x}$.}
   \label{fig:hyperfine}
\end{figure}

\autoref{fig:hyperfine} presents the root-mean-square (RMS) and maximum value of the hyperfine energy splittings calculated at each level of oxygen deficiency and with different exchange-correlation functionals. 
The trend closely follows that of the calculated magnetic moment distributions. The maximum hyperfine energy transitions range between 500-1,000 MHz with the PBE, r$^2$SCAN, and HSE06 functionals and more than 2,000 MHz with the PBE$+U$ functional. Thus, we find that some hyperfine transitions, which more broadly represent active multilevel systems (MLS), may be resonant with low frequency transmon designs based on Nb capacitors. Typical hyperfine transitions may act as a source of so-called two-level fluctuators \cite{Muller2019} with energy levels below that of the resonant frequency range of transmons but large enough to remain unsaturated at mK temperatures. Since we established that Ta$_2$O$_{5-x}$ oxides are much less likely to have local moments, this MLS mechanism is likely inactive in Ta-based qubits. 

\begin{figure}
\centering
   \includegraphics[width=0.48\textwidth]{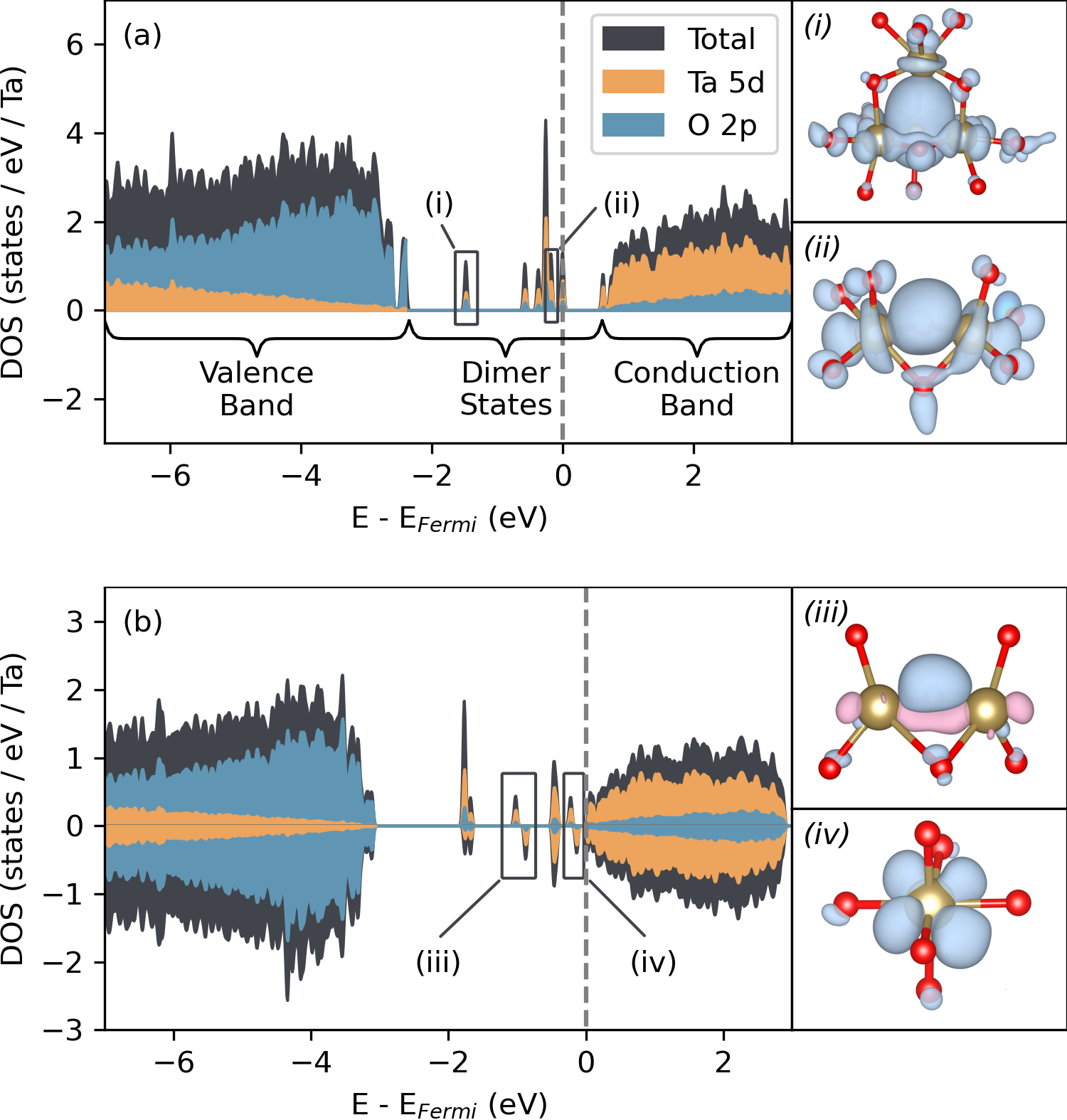} 
   \caption{(a) Total and orbital-projected density-of-states (DOS) of an amorphous Ta$_2$O$_{4.75}$ ($x=1/4$) structure with a non-magnetic ground state. (a-i, a-ii) Charge density isosurfaces of select bound states. Ta atoms are gold. O atoms are red.  (b) Total and orbital-projected DOS of an amorphous Ta$_2$O$_{4.75}$ structures with an AFM ground state. (b-iii) Spin density isosurfaces of select bound state. Spin up is blue. Spin down is pink. (b-iv) Spin density isosurface of the spin density hosted by conduction band electrons on a Ta site.} 
   \label{fig:dos}
\end{figure}

\paragraph{Metal-Metal Bonding States.}
\autoref{fig:momentstats}c-d presents the integrated average metal-metal PDFs of Nb$_2$O$_{5-x}$ and Ta$_2$O$_{5-x}$ between 0 to 3\,\AA. This window captures the nominal metal-metal dimer bond length of 2.86 \AA, while excluding other peaks, and measures how oxygen deficiencies contribute to the dimerization flaw in the amorphous structure. Both materials exhibit a nearly linear increase in their integrated PDFs as the oxygen content decreases, indicating the deficiency is accommodated in the same manner, albeit with a slightly stronger preference for formation of Ta-Ta rather than Nb-Nb dimers in the amorphous network \cite{supp}.

To understand the dimerization effect on the electronic structure and its role in magnetism, we examine the total and orbital-projected density-of-states (DOS) of one of the six Ta$_2$O$_{4.75}$ amorphous structures relaxed with the PBE functional.
\autoref{fig:dos}a shows the non-magnetic (NM) ground state has a valence band predominantly consisting of  O $2p$ states while the conduction band comprises Ta $5d$ states. Six bound states appear within the band gap. These states primarily comprise Ta $5d$ states with weak admixture of O $2p$ orbitals. To ascertain the local orbital character of these in-gap states, we computed the partial charge densities (\autoref{fig:dos}a-i,a-ii).
We find the states are $\sigma$-bonding combinations of Ta $5d$ orbitals, forming Ta-Ta dimers and trimers  with bond centered electron density. 
In some structures, tetramers are also present. Hereafter, we  refer to these $N$-mer states as ``dimer'' states for simplicity. 

\autoref{fig:dos}b shows the electronic structure of amorphous Ta$_2$O$_{4.75}$ relaxed with the PBE functional and in the AFM ground state. Similar to the NM DOS, this structure has multiple bound states. However, it has one fewer dimer state and two more electrons in the conduction band. We find that the spin density is distributed mainly in two ways \footnote{Spin density isosurfaces are rendered at an isospin density value equal to one fourth that of the isocharge density values used to render charge density isosurfaces.}: either with small local moments on the Ta sites (and some bond-centered density, \autoref{fig:dos}b-iii) or fully site-centered and localized to the crystal-field split Ta $5d$ orbitals 
(\autoref{fig:dos}b-iv). These states are then exchange split accordingly between the spin channels. 
Our simulations further reveal that the dimer states are neutral defect states analogous to $V_\mathrm{O}^0$ point defects in Ta$_2$O$_5$ and the filled conduction band states are analogous to $V_\mathrm{O}^{2+}$ defects in Ta$_2$O$_5$. 
This finding in consistent with previous work \cite{Lee2017}, which investigated the thermodynamic stability of oxygen vacancies in amorphous Ta$_2$O$_5$ and found that neutral ($V_\mathrm{O}^0$) and charged ($V_\mathrm{O}^{2+}$) defects are most stable under oxygen-rich and oxygen-poor growth conditions, respectively.
Although we did not explicitly compute the defect formation energies, our electronic structure analyses show 
that the presence of charged defect states is a necessary and sufficient condition for the formation of paramanetic moments in the oxides (excluding those relaxed with the likely unphysical PBE$+U$ simulations). 

We further identify that the underlying driving force for the formation of neutral defect states is the formation of metallic bonds between adjacent transition metal sites. Owing to the stochastic distribution of bonding environments, the formation of neutral defect states is possible in amorphous Nb$_2$O$_{5-x}$. This is in contrast to paramagnetic Nb oxide crystallographic shear structures which do not exhibit dimerization \cite{Fang2015, Kocer2019}. Nevertheless, the weaker binding energy of Nb $4d$ bonds compared to Ta $5d$ bonds is sufficient to lead to the formation of charged defect states in $>$50\% of our Nb$_2$O$_{5-x}$ structures, regardless of functional selection; whereas, charged defect states were only present in Ta$_2$O$_{5-x}$ structures relaxed with the PBE functional. These results suggest that the magnetic susceptibility of amorphous Nb$_2$O$_{5-x}$ oxides should be suppressed compared to its crystalline counterparts.

\paragraph{Local Structure--Magnetism Relationships.}
Recently, a random forest classifier model identified dependencies between preferred magnetic sites and local structural descriptors in niobium oxides \cite{Sheridan2021}. These descriptors include the minimum, mean, and maximum Nb-O bond distance, polyhedral volumes, and number of nearest neighbors. Based on our simulations, we find that magnetic moments occur either on conduction band sites or sites that participate in metal-metal dimers. In  amorphous structures, local structural features that highly influence the energies of crystal-field split transition metal $d$ orbitals in a crystalline environment,  i.e., the number and geometry of nearest neighbors and metal-oxygen bond lengths, are overshadowed by long-range disordered electrostatic interactions. Correspondingly, we do not expect conduction band sites to be highly correlated to local structural features. In contrast, metal-metal dimers that host local moments should be highly correlated to the minimum metal-metal bond distances. 

To evaluate the statistical correlations between structural features and magnetic site preference in our simulations, we computed cumulative distribution functions (CDFs) for each of the previously mentioned local structural features (except polyhedral volume) from two samples. In the first sample, we calculated the distribution of descriptors for all Nb sites. In the second sample, we calculated the distribution of descriptors for all Nb sites that hosted a local moment ($\left| \mu \right| > 0.01 \mu_B$). We then applied the two-sample Kolmogorov-Smirnov (KS) test \cite{Feller1948} to identify whether the maximum difference between the computed CDFs for each sample are sufficient to reject the null hypothesis, i.e., both CDFs arise from the same underlying distribution. See the SM for more details \cite{supp}.

We find that for our PBE simulations, the KS test fails for all structural descriptors---although only marginally for the minimum Nb-Nb distance descriptor. The KS test, however, passes in our PBE$+U$ simulations for the minimum Nb-O distance, maximum Nb-O distance, and minimum Nb-Nb distance descriptors. This is consistent with our finding that magnetic moments are hosted on conduction band sites in the  PBE simulations with only a few moments on dimer sites; in contrast, many moments are hosted on dimer sites in the PBE$+U$ simulations. Based on the correlation of several structural statistics with magnetic site preference in our PBE$+U$ simulations, cross-correlations evidently exits between Nb-O bond lengths and the minimum metal-metal distances; however, the underlying site preference is still driven by dimerization.

\paragraph{Compositional Gradients.}
While Nb$_2$O$_{5-x}$ is the dominant phase in the process-oxide layer of Nb films, the composition of the amorphous oxide layer transitions from Nb$_2$O$_5$ to Nb over a range of about one nanometer \cite{Altoe2022} where regions with nominal NbO$_2$ and NbO stoichiometry are observed. Based on our results, we expect that as the oxygen content of the amorphous oxide layer is reduced, the number of dimer states will continue to increase in  proportion to the number of oxygen atoms removed. When these bonded Nb atoms form a spanning cluster (i.e., when the percolation threshold is surpassed) the isolated Nb dimer states will form a new band of Nb $4d$ states. These states will be non-magnetic. Nevertheless, if the number of dimer states is insufficient to capture all available Nb $4d$ electrons, then paramagnetic moments will persist and may even exceed the density that we report here for Nb$_2$O$_{5-x}$ with $0\le x\le 0.25$.

\paragraph{Conclusions.} 
We employed AIMD and DFT calculations to reliably capture changes in the local (nearest neighbor distances) and medium range (octahedral connectivity) structure of  Ta$_2$O$_{5-x}$ and  Nb$_2$O$_{5-x}$. 
We found that structural flaws in the amorphous structure and tendency to local moment formation govern electron (de)localization, bound states, and activate new multilevel states that are resonant with transmon qubit frequencies.
We found that magnetic moments are fully or partially quenched in amorphous Ta$_2$O$_{5-x}$, $x\in[0,0.25]$, compared to amorphous Nb$_2$O$_{5-x}$ with equivalent oxygen concentrations. In both cases, the glass network structure accommodates the formation of predominantly neutral vacancies in the form of metal-metal dimers which do not host significant local moments. In Nb$_2$O$_{5-x}$ ($x\ge 0$), magnetic states are hosted by `conduction band' electrons associated with charged oxygen vacancies.  

Based on our results, the improved performance of qubits with either Ta or Ta capped Nb capacitors may be attributed to at least two factors: (1) The oxide layer of Ta films possesses a more robust stoichiometry compared to the oxide layer of Nb films \cite{Bal2023, Altoe2022}, which is enhanced by the increased formation energy of oxygen deficient phases. (2) For a given degree of oxygen deficiency, magnetic moment formation is suppressed in Ta$_2$O$_{5-x}$ compared to Nb$_2$O$_{5-x}$. This simultaneously suppresses losses due to the formation of Shiba states, which would introduce quasiparticle losses \cite{Sheridan2021}, and eliminates hyperfine states which may act as resonant or sub-resonant TLS. 

Our work demonstrates that while amorphous native oxides may unavoidably form during qubit fabrication with similar atomic structures, these oxides exhibit significant variations in their properties.
The superconducting components should comprise elements that lead to flexible glass networks that can capture unpaired spins to form singlet states, minimizing moment formation and potential quasiparticle or flux noise. Furthermore, because the hyperfine states form a multilevel system, which may be measured using electron paramagnetic resonance spectroscopy on these suboxides, we conjecture there may be other multilevel systems (MLS) of structural or chemical origin.
We emphasize the need for abandoning the notion of singular species alone as TLS for more complex multicomponent microscopic ensembles and devising strategies to mitigate their losses.

\begin{acknowledgments}
We would like to thank R. J. Bondi, A. P. Thompson, and M. J. Marinella for kindly providing amorphous Ta oxide structures which were used in the initial phase of this work. We would also like to thank J. F. Zasadzinski for his valuable insights and suggestions which enhanced the present work. This material is based upon work supported by the U.S. Department of Energy, Office of Science, National Quantum Information Science Research Centers, Superconducting Quantum Materials and Systems Center (SQMS) under Contract No.\ DE-AC02-07CH11359. This research used resources of the National Energy Research Scientific Computing Center, a DOE Office of Science User Facility supported by the Office of Science of the U.S.\ Department of Energy under Contract No.\ DE-AC02-05CH11231 using NERSC award BES-ERCAP0023827.
\end{acknowledgments}

\bibliography{pgpbib}

\end{document}